\documentclass[11pt]{article}

\usepackage{fullpage}

\usepackage{latexsym}
\usepackage{amsmath}
\usepackage{amssymb}
\usepackage{url}
\usepackage{xspace}
\usepackage{color}
\usepackage{graphicx}

\usepackage{booktabs}
\usepackage{listings}

\usepackage{tikz}

\usepackage{algorithm}
\usepackage{algpseudocode}

\newcommand{\leaveout}[1]{}

\newcommand{\senddata}{\textsf{Send}\xspace}
\newcommand{\recvdata}{\textsf{Recv}\xspace}

\newcommand{\bidirec}[2]{\textsf{Send}(#1)\parallel\textsf{Recv}(#2)\xspace}

\newcommand{\mpicroscope}{\textsf{mpicroscope}\xspace}

\newcommand{\mpibarrier}{\texttt{MPI\_\-Barrier}\xspace}
\newcommand{\mpiallreduce}{\texttt{MPI\_\-Allreduce}\xspace}
\newcommand{\mpireduce}{\texttt{MPI\_\-Reduce}\xspace}
\newcommand{\mpibcast}{\texttt{MPI\_\-Bcast}\xspace}
\newcommand{\mpisendrecv}{\texttt{MPI\_\-Sendrecv}\xspace}
\newcommand{\mpigetelements}{\texttt{MPI\_\-Get\_\-elements}\xspace}
\newcommand{\mpireducelocal}{\texttt{MPI\_\-Reduce\_\-local}\xspace}

\newcommand{\mpiint}{\texttt{MPI\_\-INT}\xspace}
\newcommand{\mpisum}{\texttt{MPI\_\-SUM}\xspace}

\newcommand{\openmpiversion}{Open\,MPI 4.0.5\xspace}

\newcommand{\gccversion}{\texttt{gcc 8.3.0-6}\xspace}

\title{A Doubly-pipelined, Dual-root Reduction-to-all Algorithm and Implementation}
\author{Jesper Larsson Tr\"aff\\
  TU Wien, Faculty of Informatics\\
  Treitlstrasse 3, 5th Floor/191-4, 1040 Vienna, Austria}
\date{September 28, 2021}

\begin{document}
\maketitle

\begin{abstract}
  We discuss a simple, binary tree-based algorithm for the collective
  allreduce (reduction-to-all, \mpiallreduce) operation for parallel
  systems consisting of $p$ suitably interconnected processors. The
  algorithm can be doubly pipelined to exploit
  bidirectional (telephone-like) communication capabilities of the
  communication system. In order to make the algorithm more symmetric,
  the processors are organized into two rooted trees with
  communication between the two roots. For each pipeline block, each
  non-leaf processor takes three communication steps, consisting in
  receiving and sending from and to the two children, and sending and
  receiving to and from the root.  In a round-based, uniform,
  linear-cost communication model in which simultaneously sending and
  receiving $n$ data elements takes time $\alpha+\beta n$ for system
  dependent constants $\alpha$ (communication start-up latency) and
  $\beta$ (time per element), the time for the allreduce operation on
  vectors of $m$ elements is $O(\log p+\sqrt{m\log p})+3\beta m$ by
  suitable choice of the pipeline block size. We compare the
  performance of an implementation in MPI to similar reduce followed
  by broadcast algorithms, and the native \mpiallreduce collective on
  a modern, small $36\times 32$ processor cluster. With proper choice
  of the number of pipeline blocks, it is possible to achieve better
  performance than pipelined algorithms that do not exploit
  bidirectional communication.
\end{abstract}

\section{Algorithm and Implementation}

The reduction-to-all collective communication problem is the
following. Each of $p$ successively ranked processors has a vector
$x_i$ of $m$ elements for which the element-wise product
$y=\odot_{i=0}^{p-1}x_i$ for some given associative (but not
necessarily commutative) binary operator $\odot$ has to be computed
and distributed to all $p$ processors. The effect of a
reduction-to-all operation is the same as performing the reduction of
all input vectors onto some selected root processor, say processor
$0$, and afterwards broadcasting the resulting $y$ vector to all the
other processors.

\subsection{Algorithm description}
\label{sec:algorithm}

We assume that the processors can be organized into two, roughly
equally large, rooted, post-order numbered, as balanced and complete
as possible binary trees. Processors that are in parent-child
relationship are assumed to be able to communicate directly with each
other with uniform communication costs, as are the dual
roots. Communication is bidirectional, and two parent-child processors
can in the same communication operation both send and receive elements
to and from each other (telephone-like bidirectional
communication~\cite{FraigniaudLazard94}). We assume a linear-cost
model in which bidirectional communication of $n$ elements in both
directions can be done in $\alpha+\beta n$ time units for system
dependent constants $\alpha$ (communication start-up latency) and
$\beta$ (transmission time per element).

\begin{algorithm}[t]
  \caption{Doubly pipelined reduction-to-all algorithm as performed by
    processor $i$. The constant $d_i$ is the height of processor $i$
    in its post-order numbered binary tree. Each tree root
    communicates with its dual in the other tree, and communication
    with non-existing children is void. The $j$th pipeline block is
    denoted by $Y[j]$, and blocks for which either $j<0$ or $j\geq b$
    are assumed to contain $0$ elements. The number of elements in the
    blocks $Y[j]$ and the temporary buffer $t$ that are sent and
    received is assumed to be implicitly known by the bidirectional
    send and receive operations.}
  \label{alg:doubly}
  \begin{algorithmic}[1]
    \State $Y\leftarrow x_i$ \Comment Initialize pipelining array to input $x_i$
    \For{$j=0,1,\ldots,b+d_i$}
    \State $\bidirec{Y[j-(d_i+1)],\mathrm{child}_0}{t,\mathrm{child}_0}$ \Comment First child
    \State $Y[j]\leftarrow t \odot Y[j]$ \Comment Post-order reduction
    \State $\bidirec{Y[j-(d_i+1)],\mathrm{child}_1}{t,\mathrm{child}_1}$ \Comment Second child
    \State $Y[j]\leftarrow t \odot Y[j]$ \Comment Post-order reduction
    \If {root}
    \State $\bidirec{Y[j],\mathrm{dual}}{t,\mathrm{dual}}$ \Comment Dual root
    \State $Y[j]\leftarrow t \odot Y[j]$ \Comment For non-commutative $\odot$,
    $Y[j]\odot t$ for lower numbered root
    \Else
    \State $\bidirec{Y[j],\mathrm{parent}}{Y[j-d_i],\mathrm{parent}}$
    \Comment Parent
    \EndIf
    \EndFor
  \end{algorithmic}
\end{algorithm}

The algorithm is quite simple and follows the same idea as
in~\cite{Traff06:scan} where a doubly pipelined algorithm for the
parallel-prefix operation on large vectors was discussed and
benchmarked. The input vector $x_i$ for processor $i$ is divided into
a number of $b$ successive blocks, $0<b\leq m$, that are handled one
after the other. Each processor works in a number of $b+d_i$ rounds
for some $d_i\geq 0$ to be determined later. In each such round, a
non-leaf processor (in either of the binary trees) receives a partial
result block from its first child into a temporary buffer $t$ and
bidirectionally sends an earlier block of the result $y$ to this first
child, and then performs an element-wise reduction with the
corresponding block of its own input vector. In the same way, the
processor then receives a partial result block into the temporary
buffer $t$ and sends an earlier block from and to its second child,
and performs a reduction with the received block and the partial
result block computed from the first child. In the last step, the
processor sends the partial result block to its parent, and receives
an earlier block of the result $y$ from the parent. Processors that
are leaves in their tree just send and receive blocks from their
parent processor. Thus each round entails at most three bidirectional
communication steps with roughly evenly sized blocks of about $m/b$
elements each. With the processors organized as a post-order binary
tree, the subtree rooted at some processor $i$ consists of
successively numbered processors $[i',\ldots,i'']$ and
$[i''+1,\ldots,i-1]$ for some child processors $i',i''<i$. Let the first
child of processor $i$ be processor $i-1$, and the second child be
processor $i''$. The partial result blocks computed by processor $i-1$
are blocks of the product $\odot_{k=i''+1}^{i-1}x_k$, and the partial
result blocks computed by processor $i''$ are blocks of the product
$\odot_{k=i'}^{i''}x_k$.  Thus, processor $i$ can compute partial
result blocks of the product
$\odot_{k=i'}^{i}x_k=(\odot_{k=i'}^{i''}x_k)\odot(\odot_{k=i''+1}^{i-1}x_k)\odot
x_i$ over all processors in its subtree while relying only on the
associativity of the $\odot$ operator. The root of either tree
communicates with the root of the other tree, sending partial result
blocks from its own tree and receiving partial result blocks from the
other tree. Thus, for the root an extra application of the $\odot$
operation is needed, while the other non-leaf processors take at most
two $\odot$ operations per round.

The per processor algorithm is illustrated as
Algorithm~\ref{alg:doubly}. The input for the processor is initially
stored in the $Y$ array, which will also contain the final result
$y$. The $j$th block in $Y$ is denoted $Y[j]$. Blocks for which $0\leq
j<b$ have roughly $m/b$ elements, while blocks for which $j<0$ or
$j\geq b$ are for convenience assumed to have $0$ elements. In round
$j$, processor $i$ receives block $Y[j]$ from each of its children
into an intermediate buffer $t$, and sends the previous block
$Y[j-(d_i+1)]$ of the final result. Reductions are performed on the
received blocks in $t$ with the process's own block $Y[j]$, and the
result is sent to the parent. From the parent, processor $i$ receives
block $Y[j-d_i]$ of the final result $y$. Bidirectional, simultaneous
send and receive communication is denoted by a \senddata in parallel
($\parallel$) with a \recvdata operation.

We claim that this correctly computes $y$ in $Y$ if $d_i$ is chosen as
the depth of processor $i$ in its post-ordered numbered tree. This can
easily be seen by induction on the height of the two trees. When the
height is $0$ the algorithm runs just over the two roots which in round
$j, 0\leq j<b$ just exchanges their input blocks $Y[j]$ and compute
the result correctly into $Y[j]$ by a single application of
$\odot$. Assume the claim holds for any two binary trees of height at
most $d, d\geq 0$, and consider a processor of depth $d$ with one or
two children at depth $d+1$. In round $j$ such a processor will need
to compute a partial result in $Y[j]$ to send to its parent in the
third step. For this it needs to receive $Y[j]$ from its children,
which is what the algorithm does. Before round $j$ such a processor
has by the induction hypothesis correctly received all result blocks
of $y$ in $Y[-d,\ldots, j-1-d]$. In the first two steps, it sends the
last of these blocks, namely $Y[j-1-d]=Y[j-(d+1)]$ to its children,
and in the third step receives block $Y[j-d]$ of the result $y$.

\subsection{Analysis remarks}

Assume that for the number of processors $p$, it holds that $p+2=2^h$
for some $h, h>0$, that is $p=2^h-2$, and that $h>1$.  Then, the
height of the two binary trees is $h-1$, and the number of
communication rounds for the first block to reach a tree root is
$2(h-1)$ (as can easily be shown by induction on $h$). Each root
receives a new block every third step since the algorithm takes three
send and receive operations per round. One extra step is required for
either root to receive the first block from the other tree. For the
last block that is broadcast down the tree, another $2(h-1)$ steps is
needed. Thus the latency of the doubly pipelined algorithm in terms of
the number of communication steps for the first block of the result
$y$ to reach the last leaf of either of the binary trees is
$2(h-1)+1+2(h-1)=4h-3$. Each subsequent block requires three steps.

Assuming a linear-cost communication model with known constants
$\alpha$ and $\beta$, the time to perform the allreduce operations on
the $m$ element vectors when divided into $b$ blocks of roughly $m/b$
elements each is thus
\begin{displaymath}
  (4h-3+3(b-1))(\alpha+\beta m/b) \quad .
\end{displaymath}
By balancing (``Pipelining Lemma'') the terms that increase and decrease
with $b$, the analytical best number of blocks and with this the best
possible running time of
\begin{displaymath}
  (4h-6)\alpha + 2\sqrt{3(4h-6)\alpha\beta m} + 3\beta m \quad
\end{displaymath}
can easily be deduced, which is $O(\log p+\sqrt{m\log p})+3\beta m$.

The analysis accounts only for the communication costs. All non-leaves
except the roots perform two applications of the $\odot$ operator on
blocks of $m/b$ elements per round. The two roots unfortunately need
one more $\odot$ reduction with the partial result block received from
the other root. With a cost of $\gamma$ time units per element, the
added cost for the reductions is thus at most $3 \gamma m/b$ per
round.

If instead only one, doubly pipelined binary tree is used, all
non-leaves, including the root, perform at most two applications of
the $\odot$ operator per round. On the other hand, with only one
binary tree and $p+1=2^h$, the latency for the first block of the
result $y$ to reach the last child is $4h$, and therefore slightly
higher (by a small constant term).

If the reduction-to-all operation is implemented as a reduction
operation to the root followed by a broadcast, both with binary
trees, the total time is
\begin{displaymath}
  2(2h+2(b-1))(\alpha +\beta m/b)
\end{displaymath}
which results in a running time of $O(\log p+\sqrt{m\log p})+4\beta m$
with the right choice of best number of blocks.  Thus, in the
$\beta$-term, exploiting bidirectional communication with doubly
pipelined trees gives an improvement from a factor $4$ to a factor of
$3$.

The best-known, pipelined, binary tree-based algorithm places the $p$
processors in two trees, such that each processor is an internal node
in one tree and a root in the other~\cite{Traff09:twotree}. This gives
a running time for the reduction-to-all operations, when implemented
as a reduction followed by a broadcast operation of $O(\log
p+\sqrt{m\log p})+2\beta m$.

\subsection{Implementation sketch}
\label{sec:implementation}

A concrete implementation of the algorithm has been given using
MPI~\cite{MPI-3.1}. The bidirectional communication is assumed to be
effected with the \mpisendrecv operation. When a processor has not yet
partially reduced all blocks, a block of size $b$ is received from
each of the children and sent to the parent; when all blocks have been
reduced, virtual blocks of zero elements are received and
sent. Likewise, blocks of zero elements are initially received from
the parent and sent to the children, and first when the parent has
received blocks with more than zero elements, these are transmitted to
the children. For the implementation, each \mpisendrecv operation
gives the upper bound, namely $b$ elements, on the size of the block
expected to be received, and the actual number of elements in a
received block is queried with \mpigetelements. Using this
functionality, there is no need to explicitly keep track of the depth
of the processor and the excess number of rounds $d$. A non-leaf
processor terminates when it has received its last non-zero element
block from both its children and the parent, but since the blocks
received from the parent are always behind (earlier than) blocks from
the children, a processor can terminate as soon as it has received the
last non-zero element block from the parent. The \mpireducelocal
function is used for performing the block wise reductions; but since
this is less flexible than the assignments shown in
Algorithm~\ref{alg:doubly}, some care has to be taken to respect the
possible non-commutativity of the operator, and to avoid extra buffer
copying.

The whole algorithm can be implemented in less than hundred of lines of
MPI C code. Such code is available from the author.

\begin{table}
  \caption{Systems (hardware and software) used for the experimental
    evaluation.
  }
  \label{tab:systems}
  \begin{center}
    \begin{tabular}{crrrccc}
      Name & $n$ & $N$ & $p$ & Processor & Interconnect & MPI library \\
      \toprule
      Hydra & 32 & 36 & 1152 & Intel Xeon Gold 6130, 2.1 GHz & OmniPath & \openmpiversion \\
      & & & & & & with \gccversion \\
      & & & & Dual socket & Dual (2-lane) & \\ 
      \bottomrule
    \end{tabular}
  \end{center}
\end{table}

\section{An experimental evaluation}

We have done an initial evaluation of the doubly pipelined, dual-root
reduction-to-all implementation on a small, Intel Skylake dual-socket,
dual-rail OmniPath ``Hydra'' cluster as described in
Table~\ref{tab:systems}. The nodes of this cluster consists of two
16-core sockets, each with a direct OmniPath connection to a separate
network.

The evaluation compares the following implementations of the
reduction-to-all operation on vectors of given numbers of elements.

\begin{enumerate}
  \item
    The native \mpiallreduce operation.
  \item
   An \mpireduce followed by an \mpibcast operation.
\item
  A pipelined reduce followed by a pipelined broadcast with the same
  pipeline block size using a single binary tree (\textsf{User-Allreduce1}).
\item
  The doubly pipelined, dual root reduction-to-all algorithm
  implemented as sketched in Sections~\ref{sec:algorithm}
  and~\ref{sec:implementation} (\textsf{User-Allreduce2}).
\end{enumerate}
The two implemented pipelined algorithms use the same pipeline block
size which is set at compile time. The implementations do not attempt
to find and use a best block size in dependence on the number of
elements to reduce and the number of MPI processes used, or other
characteristics of the system. Experiments with different block sizes,
different numbers of MPI processes, and different mappings of the MPI
processes to the cores of the compute nodes must be performed.

\begin{figure}
  \includegraphics[width=0.48\textwidth]{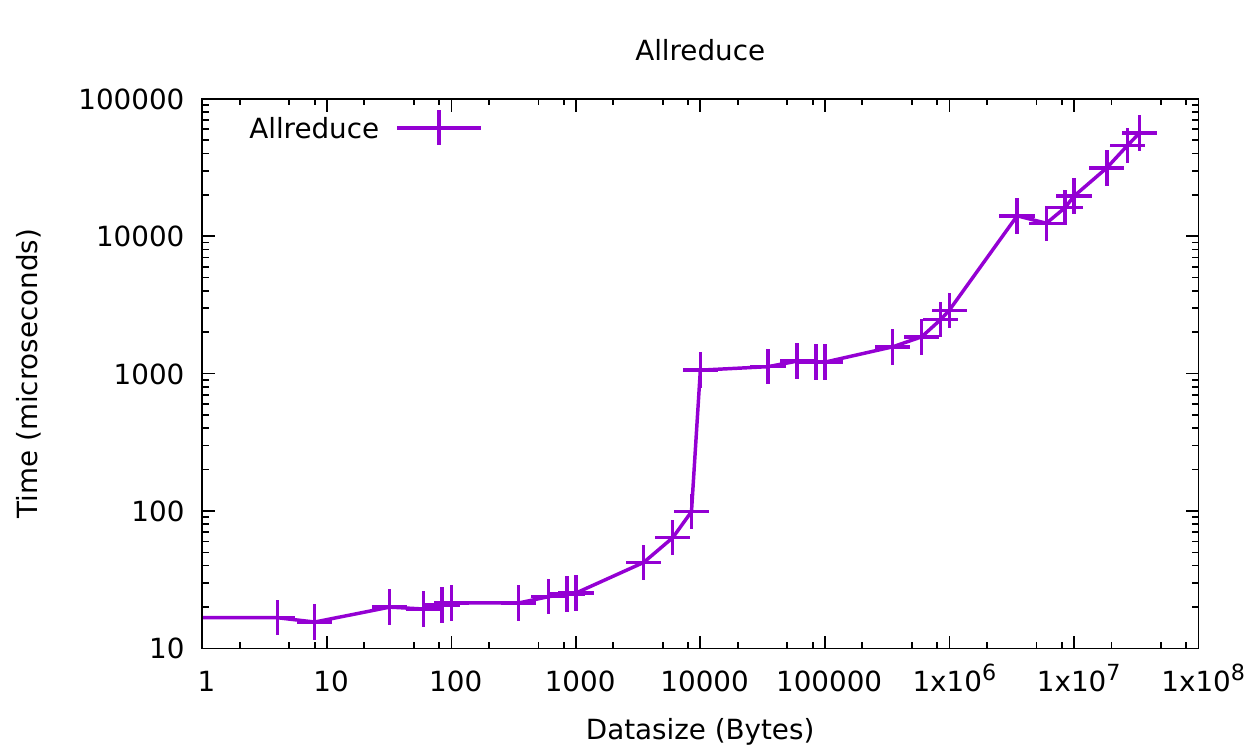}
  \includegraphics[width=0.48\textwidth]{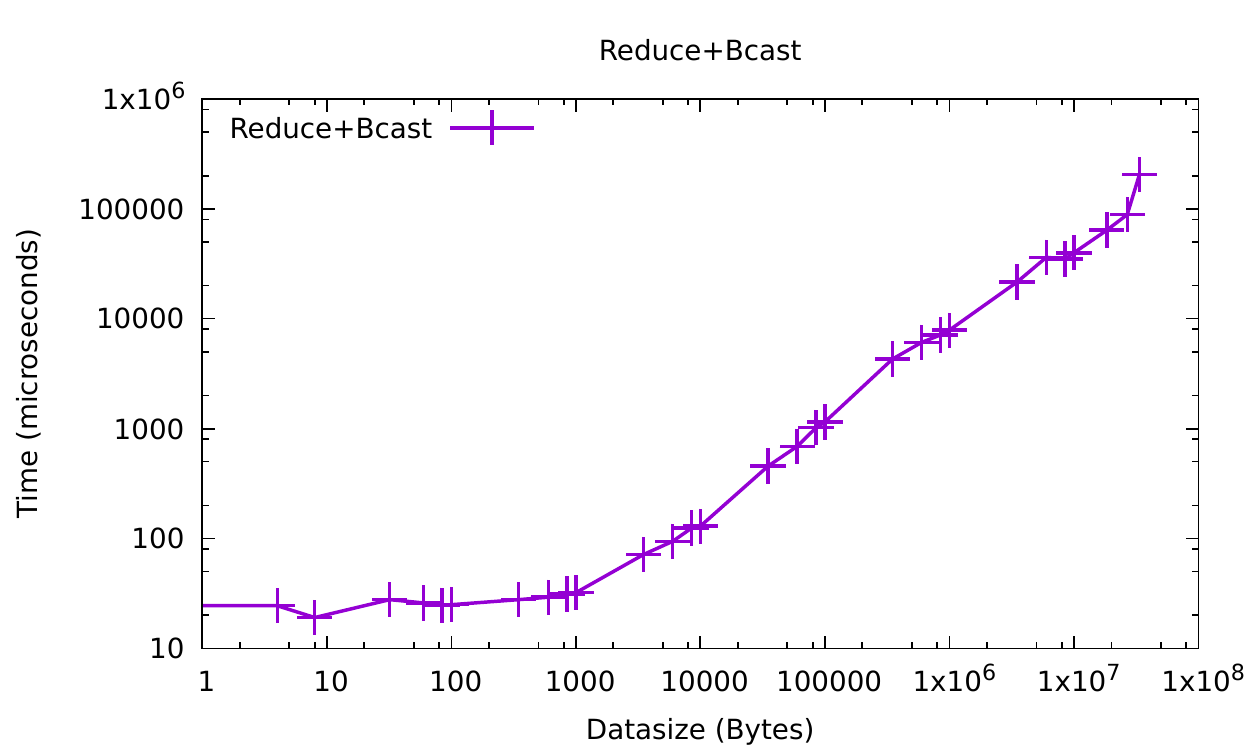}
  \\
  \includegraphics[width=0.48\textwidth]{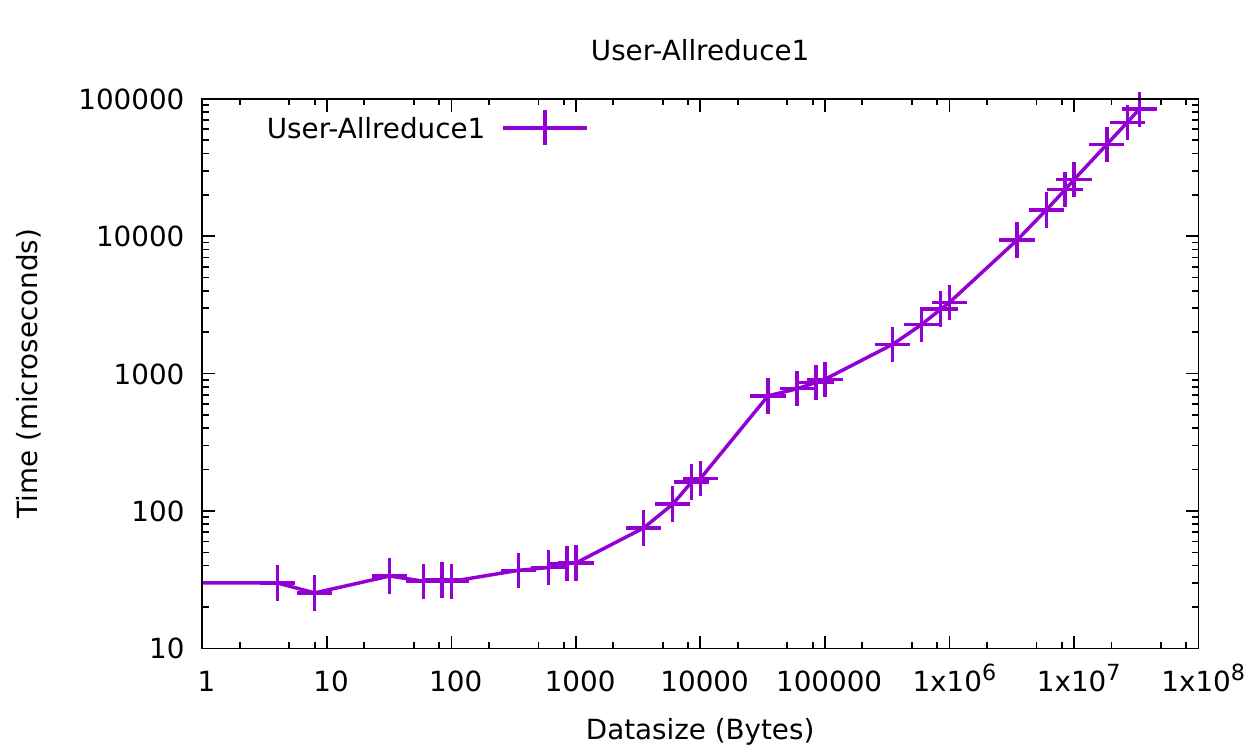}
  \includegraphics[width=0.48\textwidth]{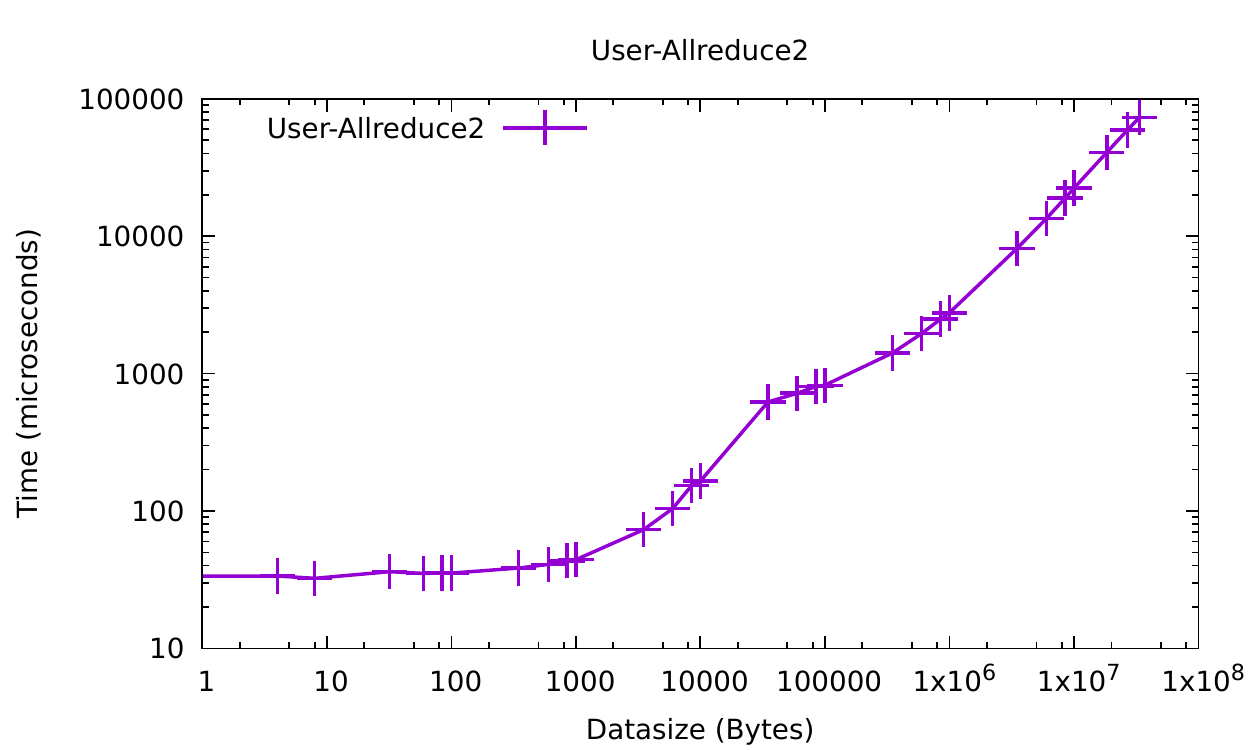}
  \caption{Runs with the four different reduction-to-all
    implementations. The two pipelined algorithms pipeline with
    $b=16000$ blocks of \mpiint elements. \textsf{User-Allreduce1}
    implements the pipelined reduction followed by broadcast
    algorithm. \textsf{User-Allreduce2} implements the doubly
    pipelined, dual root algorithm.}
\label{fig:result}
\end{figure}

This is all not done here (see Section~\ref{sec:summary} for
justification). In Figure~\ref{fig:result} results from a single run
with $p=36\times 8$ MPI processes (8 processes on each of the 36
compute nodes) with a fixed number of pipeline blocks of $b=16000$
elements is shown. The elements are \mpiint integers, and the
reduction operator used in \mpisum. The results are gathered using the
\mpicroscope benchmark~\cite{Traff12:mpibenchmark}. This benchmark
defines the running time of an experiment as the minimum over a number
of measurement rounds of the completion time of the slowest MPI
process, and synchronizes individual measurements with \mpibarrier
operations~\cite{HunoldCarpenamarie16}.

\begin{table}
  \caption{Raw data for the experiment with the four algorithms with
    $b=16000$ pipeline blocks. The minimum completion times over a
    number of measurements are shown.}
  \label{tab:rawdata}
  \begin{center}
    \begin{tabular}{crrrr}
      Elements (count) & \mpiallreduce & \texttt{MPI\_Reduce}+\mpibcast &
      Pipelined & Doubly pipelined \\
      \toprule
0	&	0.29	&0.84	&0.19	&0.19	\\
1	&	16.75	&24.44	&29.95	&33.60	\\
2	&	15.55	&19.08	&25.37	&32.41	\\
8	&	19.96	&27.80	&33.60	&36.18	\\
15	&	19.38	&25.91	&30.82	&35.12	\\
21	&	20.63	&24.64	&31.45	&35.49	\\
25	&	21.45	&24.98	&30.70	&35.32	\\
87	&	21.46	&27.82	&37.08	&38.48	\\
150	&	23.77	&29.23	&38.84	&40.88	\\
212	&	24.96	&31.21	&41.32	&43.41	\\
250	&	25.27	&32.17	&41.82	&44.25	\\
875	&	42.19	&71.44	&75.20	&73.15	\\
1500	&	63.98	&94.17	&112.31	&104.39	\\
2125	&	99.07	&124.31	&162.24	&152.74	\\
2500	&	1059.83	&129.21	&172.16	&165.21	\\
8750	&	1122.91	&456.07	&689.72	&621.82	\\
15000	&	1233.48	&688.99	&775.91	&719.73	\\
21250	&	1218.00	&1020.46&	862.24	&805.53	\\
25000	&	1211.81	&1146.03&	908.35	&822.63	\\
87500	&	1563.37	&4294.96&	1630.25	&1412.93\\	
150000	&	1854.84	&6087.61&	2276.29	&1958.36\\	
212500	&	2472.61	&7106.53&	2941.19	&2489.45\\	
250000	&	2893.00	&7835.16&	3289.41	&2765.93\\	
875000	&	14083.86&21566.69&9392.92	&8158.38\\	
1500000	&	12421.02&36192.82&15557.71&13434.51\\
2125000	&	16154.38&34915.25&21776.97&18955.76\\
2500000	&	19579.38&39681.02&25773.33&22346.98\\
4597152	&	31391.74&63723.56&46497.68&40701.29\\
6694304	&	45622.58&88317.08&67372.14&59036.27\\
8388608	&	56249.24&204326.0&84081.41&73116.03\\
      \bottomrule
    \end{tabular}
  \end{center}
\end{table}

The raw data are listed in Table~\ref{tab:rawdata}. The data range is
from $0$ to $40\,000\,000$ Bytes with exponentially distributed measure
points as chosen by the \mpicroscope benchmark. 

As can be seen, the doubly pipelined algorithm consistently (except
for small counts) beats the pipelined reduction followed by broadcast
algorithm, but the ratio of improvement in time is less than the
factor of $4/3$ as expected by the analysis (for instance, for the
largest count, the ratio is only $1.14$ and not $1.33$), which may or
may not indicate that bidirectional communication capabilities are
being exploited. In order to answer this question, a baseline on raw,
bidirectional communication would need to be experimentally
established. For small and large counts, the native \mpiallreduce
operation performs the best, but is excessively poor in a midrange of
counts, where it is the worst implementation by a sometimes large
factor. This indicates a bad switch of algorithm in the used
\openmpiversion library. As counts get larger, the poorest
implementation choice is \texttt{MPI\_Reduce}+\mpibcast, as is the way
an MPI library can be expected to behave~\cite{Traff10:selfcons}.

\section{Summary}
\label{sec:summary}

This note is meant as an exercise in reduction-algorithm
implementation and evaluation, and most concrete issues are therefore
intentionally left open. The main question is whether bidirectional
communication in message-passing systems can be exploited and make a
noticeable and robust performance difference over algorithms that
cannot exploit bidirectional communication. Further questions
concern the experimental evaluation, in particular the determination
of the best pipeline block size, and the role of the hierarchical
structure (network and nodes) of a clustered, high-performance system
play. The invitation and challenge is to investigate these questions
better than presented here. Concrete implementations can be compared
against the author's code.

\bibliographystyle{plain}
\bibliography{traff,parallel}

\end{document}